\documentclass[epjc3]{svjour3} 

\usepackage[margin=25mm]{geometry}

\usepackage{amsmath,amssymb}

\usepackage{cite}

\usepackage{url}

\def\<#1>{\langle #1 \rangle_{q}}
\newcommand{\ExpectME}[1]{\langle #1 \rangle_{q}^{\mathrm{ME}}} 
\newcommand{\ExpectMEwithEVN}[1]{\langle #1 \rangle_{q,(EVN)}^{\mathrm{ME}}} 
\newcommand{\ExpectCE}[1]{\langle #1 \rangle_{q}^{\mathrm{CE}}}
\newcommand{\bra}[1]{\langle #1|}
\newcommand{\ket}[1]{| #1 \rangle}
\newcommand{\SR}{S_{\mathrm{R}q}}
\newcommand{\ST}{S_{\mathrm{T}q}}
\newcommand{\SRME}{S^{\mathrm{ME}}_{\mathrm{R}q}}
\newcommand{\STME}{S^{\mathrm{ME}}_{\mathrm{T}q}}
\newcommand{\SRCE}{S^{\mathrm{CE}}_{\mathrm{R}q}}
\newcommand{\STCE}{S^{\mathrm{CE}}_{\mathrm{T}q}}
\newcommand{\One}{\hat{1}}
\newcommand{\hA}{\hat{A}}
\newcommand{\hH}{\hat{H}}
\newcommand{\Tr}[1]{\mathrm{Tr}\left[#1\right]}
\newcommand{\TrDelta}[1]{\mathrm{Tr}_{\Delta}\left[#1\right]}  

\newcommand{\qmrho}{\hat{\rho}}
\newcommand{\qmrhoME}{\hat{\rho}^{\mathrm{ME}}}
\newcommand{\qmrhoCE}{\hat{\rho}^{\mathrm{CE}}}
\newcommand{\Tph}{T_{\mathrm{ph}}}
\newcommand{\TphME}{T_{\mathrm{ph}}^{\mathrm{ME}}}
\newcommand{\TphCE}{T_{\mathrm{ph}}^{\mathrm{CE}}}

\newcommand{\betaCE}{\beta^{\mathrm{CE}}}
\newcommand{\betaPhME}{\beta_{\mathrm{ph}}^{\mathrm{ME}}}
\newcommand{\betaPhCE}{\beta_{\mathrm{ph}}^{\mathrm{CE}}}
\newcommand{\CvCE}{C_V^{\mathrm{CE}}}
\newcommand{\CvME}{C_V^{\mathrm{ME}}}

\newcommand{\DE}{\Delta E}%

\title{Relation between the escort average in microcanonical ensemble and the escort average in canonical ensemble in the Tsallis statistics}
\author{Masamichi Ishihara\thanksref{e1,addr1}}
\thankstext{e1}{email: m\_isihar@koriyama-kgc.ac.jp}
\institute{Department of Food and Nutrition, Koriyama Women's University, Koriyama, Fukushima, 963-8503, Japan \label{addr1}}

\begin{document}

\abstractdc{
We studied the escort averages in microcanonical and canonical ensembles in the Tsallis statistics of entropic parameter $q>1$.
The quantity $(q-1)$ is the measure of the deviation from the Boltzmann-Gibbs statistics.
We derived the relation between the escort average in the microcanonical ensemble and the escort average in the canonical ensemble.
Conditions arise by requiring that the integrals appeared in the canonical ensemble do not diverge.
A condition is the relation between the heat capacity $\CvCE$ at constant volume in the canonical ensemble and the entropic parameter $q$: $0 < (q-1) \CvCE < 1$.
This condition gives the known condition when $\CvCE$ equals the number of ingredients $N$. 
With the derived relation, 
we calculated the energy, the energy fluctuation, 
and the difference between the canonical ensemble and the microcanonical ensemble in the expectation value of the square of Hamiltonian.
The difference between the microcanonical ensemble and the canonical ensemble in energy is small because of the condition.
The heat capacity $\CvCE$ and the quantity $(q-1)$ are related to the energy fluctuation and the difference.
It was shown that the magnitude of the relative difference $|(\SRCE-\SRME)/\SRME|$ is small when the number of free particles is large,  
where $\SRME$ is the R\'enyi entropy in the microcanonical ensemble and $\SRCE$ is the R\'enyi entropy in the canonical ensemble.
The similar result was also obtained for the Tsallis entropy.
}

\maketitle

\section{Introduction}
The statistical mechanics is widely used to describe a large number of phenomena, 
and the extension of the standard statistics, Boltzmann-Gibbs statistics, has been attempted.
A candidate is the Tsallis statistics which shows power-like distributions. 
This statistics has been applied in various branches of science \cite{Tsallis:Book}, 
such as isotropic spin-1/2 $XX$ dimer \cite{Castano:PRE104:2021}, high energy heavy ion collisions \cite{Saha:MPLA36:2021}, etc.

The Tsallis statistics is based on the Tsallis entropy. 
The R\'enyi entropy resembles the Tsallis entropy, and the R\'enyi entropy can be obtained by transforming the Tsallis entropy. 
In this statistics, the escort average is often used to calculate quantities.
Therefore, the escort average of Hamiltonian is used for the energy constraint.
The probability or density operator can be obtained by applying the maximum entropy principle
\cite{Jaynes:1957,Tsallis:Book,Tsallis:PhysicaA:1998,Ishihara:2020:QuantAnal:DensityOp,Ishihara:2021:QnantAnal:GGE}.
The optimal Lagrange multipliers method \cite{Martinez:PhysicaA286:2000,Shen:PhysicaA487:2017} 
is an method of the maximum entropy principle, 
and is also applied to obtain the probability in the Tsallis statistics with escort average. 

The Lagrange multiplier corresponding to the standard energy constraint is interpreted as the inverse of the temperature in the Tsallis statistics, 
where we call the temperature the Tsallis temperature to avoid confusion.
Another temperature, so-called the physical temperature
\cite{Kalyana:2000,Abe-PLA:2001,S.Abe:physicaA:2001,Aragao:2003,Ruthotto:2003,Toral:2003,Suyari:2006,Ishihara:phi4,Ishihara:free-field,Ishihara:Thermodyn-rel},
was introduced as a variable to describe the equilibrium of subsystems.
It may be better to call the temperature the equilibrium temperature \cite{Ishihara:Thermodyn-rel,Imdiker_2023}.
The inverse of the Tsallis temperature is given by the derivative of the Tsallis entropy with respect to the energy,
and the inverse of the physical temperature is given by the derivative of the R\'enyi entropy with respect to the energy.
Therefore, the physical temperature is related to the Tsallis temperature.

Some ensembles are used to calculate physical quantities in statistical mechanics.
The relation between the quantity in the microcanonical ensemble and the quantity in the canonical ensemble is known \cite{Saso:Book:2010}.
For example, the expectation value of the squared energy in the canonical ensemble is not equal to that in the microcanonical ensemble,
because the energy fluctuates \cite{Vives2002,Liyan2008} in the canonical ensemble.
The large differences might arise because a distribution in the Tsallis statistics has a long tail.
It is worth to study the system in different ensembles in the Tsallis statistics \cite{Ishihara:2023}.

In this paper, we attempt to find the relation between the quantity in the microcanonical ensemble and the quantity in the canonical ensemble
in the Tsallis statistics of entropic parameter $q>1$
by assuming that the physical temperature in the microcanonical ensemble is equal to the physical temperature in the canonical ensemble.
The quantity $(q-1)$ is the measure of the deviation from the Boltzmann-Gibbs statistics.
It may be worth to mention that the value of $q$ is often restricted \cite{Wilk2000,Ishihara2015,Ishihara:2016,Bhattacharyya:2016,Bhattacharyya:2018}.
With the derived relation, we study the relations for the energy,
the expectation value of $\hH^2$, the R\'enyi entropy, and the Tsallis entropy, where $\hH$ is the Hamiltonian.

We derived the relation between the escort average in the microcanonical ensemble and the escort average in the canonical ensemble,
and found the condition which is the relation between the heat capacity $\CvCE$ at constant volume in the canonical ensemble 
and the measure $(q-1)$: $0 < (q-1) \CvCE < 1$. 
With the derived relation,
we obtained the difference between the energy in the canonical ensemble and the energy in the microcanonical ensemble.
We showed that the energy fluctuation in the canonical ensemble is related to $\CvCE$ and $(q-1)$.
We also showed that
the difference between the microcanonical ensemble and the canonical ensemble
in the expectation value of $\hH^2$ is related to $\CvCE$ and $(q-1)$.
Additionally, we showed that the magnitude of the relative difference $|(\SRCE-\SRME)/\SRME|$ is small for the large number of free particles,
where $\SRME$ is the R\'enyi entropy in the microcanonical ensemble and $\SRCE$ is the R\'enyi entropy in the canonical ensemble.
The similar result was also shown for the Tsallis entropy.

This paper is organized as follows.
In Sec.~\ref{sec:review}, we review the Tsallis statistics briefly. 
In Sec.~\ref{sec:derive-relation},
we derive the relation between the escort average in the microcanonical ensemble and the escort average in the canonical ensemble,
and also calculate the partition function. 
We derive the condition between the measure $(q-1)$ and the heat capacity at constant volume from these calculations.
In Sec.~\ref{sec:energy-fluctuation-entropies},
we apply the derived relation to $\hH$ and $\hH^2$, 
and calculate the energy, the energy fluctuation, and 
the difference between the microcanonical ensemble and the canonical ensemble in the expectation value of $\hH^2$. 
We calculate the magnitude of the relative difference between the microcanonical ensemble and the canonical ensemble for the R\'enyi entropy, 
and we also calculate that for the Tsallis entropy.
The last section is assigned for discussions and conclusions.

\section{Brief review of the Tsallis statistics}
\label{sec:review}
The Tsallis statistics is based on the Tsallis entropy defined by 
\begin{align}
  \ST &= - \Tr{\qmrho^q \ln_q (\qmrho)} = \frac{\Tr{\qmrho^q}-1}{1-q}, 
\end{align}
where $\qmrho$ is the density operator, $q$ is the entropic parameter, 
and $\ln_q (x)$ represents the $q$-logarithmic function. 
The escort average is often used as the expectation value.
The escort average of a quantity $\hA$ is defined by
\begin{align}
\< \hA > = \frac{\Tr{\qmrho^q \hA}}{\Tr{\qmrho^q}}.
\end{align}

The density operator in the microcanonical ensemble is derived by applying the maximum entropy principle
with the normalization condition $\Tr{\qmrho} = 1$.
The resulting Tsallis entropy is
\begin{align}
  \STME = \ln_q W,  
\end{align}
where $W$ is the number of state. 
The density operator in the canonical ensemble is derived by applying the maximum entropy principle
with the normalization condition $\Tr{\qmrho} = 1$ and the energy constraint $\<\hH> = U$: 
\begin{subequations}
\begin{align}
  &\qmrhoCE = \frac{1}{Z} \exp_q\left(-\frac{\beta}{c_q} (\hH - U)\right), \label{canonical:rho}\\
  &Z = \Tr{\exp_q\left(-\frac{\beta}{c_q} (\hH - U)\right)}, 
\end{align}
\end{subequations}
where $\hH$ is the Hamiltonian, $\exp_q(x)$ is the $q$-exponential function, $c_q$ is $\Tr{(\qmrhoCE)^q}$, and $Z$ is the partition function.
It is noted that the partition function $Z$ contains the energy $U$.

The Tsallis entropy in the canonical ensemble is represented as 
\begin{align}
&\STCE = \frac{\Tr{(\qmrhoCE)^q}-1}{1-q}. 
\end{align}
The R\'enyi entropy $\SR$ and Tsallis entropy $\ST$ are related each other: 
\begin{align}
\SR = \frac{1}{1-q} \ln (1+(1-q)\ST).  
\end{align}

The physical temperature, which characterizes the equilibrium, was introduced \cite{Abe-PLA:2001,S.Abe:physicaA:2001}.
The Tsallis entropy has peudo-additivity: $\ST^{A+B} = \ST^A + \ST^B + (1-q) \ST^A \ST^B$
for the subsystem $A$ with the energy $E^A$ and the subsystem $B$ with the energy $E^B$.
The maximum entropy principle requires $\delta \ST^{A+B} = 0$ and we have
\begin{align}
\frac{\delta \ST^A}{1+(1-q) \ST^A} = - \frac{\delta \ST^B}{1+(1-q) \ST^B}.
\end{align}
With the energy conservation $\delta (E^A+E^B) = 0$, we have
\begin{align}
\frac{1}{1+(1-q) \ST^A} \frac{\delta \ST^A}{\delta E^A} = \frac{1}{1+(1-q) \ST^B} \frac{\delta \ST^B}{\delta E^B}.
\end{align}
Therefore, the physical temperature $\Tph$ of the system with the energy $E$ is defined by 
\begin{align}
\frac{1}{\Tph} = \frac{1}{1+(1-q) \ST} \frac{\partial \ST}{\partial E} = \frac{\partial \SR}{\partial E}.
\end{align}

We attempt to find the relation between
the escort average in the microcanonical ensemble and the escort average in the canonical ensemble 
in Sec.~\ref{sec:derive-relation}.

\section{The relation between the escort average in the microcanonical ensemble and that in the canonical ensemble and the calculation of the partition function}
\label{sec:derive-relation}
We employ the Tsallis entropy and the escort average in this paper.  
In this section, we attempt to derive the relation
between the escort average in the microcanonical ensemble and the escort average in the canonical ensemble,
using the standard method \cite{Saso:Book:2010}.
We also attempt to calculate the partition function.

\subsection{The number of states and the density of states}
We use the standard procedure in statistical mechanics. 
We introduce the number of states $W(E,V,N,\DE)$ for the volume $V$ and the number of ingredients $N$ 
in the range of $[E-\DE/2,E+\DE/2)$, where $E$ is the energy and $\DE$ is the energy width.
We also introduce the number of states $\Omega_0(E,V,N)$ below $E$ and the density $\Omega(E,V,N)$ given by
\begin{align}
\Omega(E,V,N) = \frac{\partial \Omega_0(E,V,N)}{\partial E} .
\end{align}

The quantity $W$ is expanded with respect to $\DE$.
\begin{align}
  W(E,V,N,\DE) &= W(E,V,N,\DE=0) + \left. \frac{\partial W(E,V,N,\DE)}{\partial (\DE)} \right|_{\DE=0} \DE + O((\DE)^2) \nonumber \\
  &= \left. \frac{\partial W(E,V,N,\DE)}{\partial (\DE)} \right|_{\DE=0} \DE + O((\DE)^2) .
\label{eqn:W:diff-W}
\end{align}
The quantity $\left. \partial W(E,V,N,\DE)/\partial (\DE) \right|_{\DE=0}$ is larger than or equal to zero.

It is possible to represent $\Omega_0$ with $W$. Let $M$ be positive integer with $\DE = E/M$.
We have
\begin{align}
\Omega_0(E,V,N) - \Omega_0(E=0,V,N) = \sum_{j=0}^{M-1} \left. W(E, V, N, \Delta E) \right|_{E=E_j}, \qquad E_j = (j+1/2)\DE .
\end{align}
From Eq.~\eqref{eqn:W:diff-W}, $\Omega_0$ is represented for sufficiently large $M$:
\begin{align}
  \Omega_0(E,V,N) - \Omega_0(E=0,V,N)
  &\sim \sum_{j=0}^{M-1} \left[ \left. \frac{\partial W(E,V,N,\DE)}{\partial (\DE)} \right|_{\DE=0} \right]_{E=E_j} \DE
  \nonumber \\ 
  &\sim \int_0^{E} dE' \left.\frac{\partial W(E',V,N,\DE)}{\partial (\DE)} \right|_{\DE=0} .
\end{align}
Therefore, we obtain
\begin{align}
  \Omega(E,V,N) = \left.\frac{\partial W(E,V,N,\DE)}{\partial (\DE)} \right|_{\DE=0} .
\end{align}
We use this relation in the next subsection.

\subsection{The relation between the escort average in the microcanonical ensemble and the escort average in the canonical ensemble}
The maximum entropy principle is often used to obtain the density operator. 
The functional $I^{\mathrm{ME}}$ in the microcanoncial ensemble is defined by
\begin{align}
  I^{\mathrm{ME}} = S - \alpha (\TrDelta{\qmrho} - 1), 
\end{align}
where $S$ is the entropy, $\alpha$ is the Lagrange multiplier, 
and $\mathrm{Tr}_{\Delta}$ is the trace for the states in the range $[E-\DE/2,E+\DE/2)$. 
The maximum entropy principle requires $\delta I^{\mathrm{ME}} = 0$.

The density operator in the microcanonical ensemble for the Tsallis entropy is given by 
\begin{align}
\qmrhoME = \frac{\One}{W(E,V,N,\DE)} . 
\end{align}
Therefore, we have the Tsallis entropy and the R\'enyi entropy in the microcanonical ensemble \cite{Abe:PRE66:2002,Moyano:EuroLett73:2006}: 
\begin{subequations}
\begin{align}
\STME &= \ln_q W(E,V,N,\DE), \\
\SRME &= \ln W(E,V,N,\DE) .
\end{align}
\end{subequations}
The physical temperature $\TphME$ in the microcanonical ensemble is given by
\begin{align}
\frac{1}{\TphME(E)}  = \frac{\partial \SRME}{\partial E}. 
\end{align}

The escort average of a quantity $\hA$ in the microcanonical ensemble is 
\begin{align}
  \ExpectME{\hA} = \frac{\TrDelta{\left(\qmrhoME\right)^q \hA}}{\TrDelta{\left(\qmrhoME\right)^q}}
  = \frac{1}{W(E,V,N,\DE)} \sum_{j=1}^{W(E,V,N,\DE)} \bra{j} \hA \ket{j},
  \quad
  E_j \in [E-\DE/2, E+\DE/2),
\end{align}
where $\ket{j}$ is the state with the energy $E_j$ labeled $j$. 
The well-known expression of the expectation value $\ExpectME{\hA}$ for small $\DE$ is obtained by using the delta-function:
\begin{align}
  \ExpectME{\hA}
  = \frac{\DE}{W(E,V,N,\DE)} \sum_{j=1}^{W(E,V,N,\DE)} \delta(E-E_j) \bra{j} \hA \ket{j} 
  &  \sim \frac{1}{\Omega(E,V,N)} \sum_{j=1}^{W(E,V,N,\DE)} \delta(E-E_j) \bra{j} \hA \ket{j} .
\end{align}

In the canonical ensemble, the following functional is extremized:
\begin{align}
  I^{\mathrm{CE}} = S - \alpha (\Tr{\qmrho} - 1) - \betaCE (\ExpectCE{\hH} - U ), 
\end{align}
where $\mathrm{Tr}$ means the trace, and $\alpha$ and $\betaCE$ are the Lagrange multipliers. 
The expectation value $\ExpectCE{\hA}$ is the escort average of $\hA$ in the canonical ensemble: 
\begin{align}
  \ExpectCE{\hA} &= \frac{\Tr{\left(\qmrhoCE\right)^q \hA}}{\Tr{\left(\qmrhoCE\right)^q}}.
\end{align}

For the Tsallis entropy, the density operator in the canonical ensemble \cite{Tsallis:PhysicaA:1998,Suyari:2006} is obtained:
\begin{subequations}
\begin{align}
  \qmrhoCE &= \frac{1}{Z} \exp_q( -\betaPhCE \cdot (\hH-\ExpectCE{\hH})),\\
  & Z = \Tr{\exp_q( -\betaPhCE  \cdot  (\hH-\ExpectCE{\hH}))}, \label{eqn:CE:partition-function}\\
  & \betaPhCE = \betaCE/\Tr{(\qmrhoCE)^q}.
\end{align}
\end{subequations}
With the relation $U=\ExpectCE{\hH}$, we have
\begin{align}
  \ExpectCE{\hA}(U)
  = \frac{\displaystyle\sum_n [\exp_q(-\betaPhCE  \cdot (E_n-U))]^q  \bra{n} \hA\ket{n}}
  {\displaystyle\sum_n [\exp_q(-\betaPhCE  \cdot  (E_n-U))]^q},
\end{align}
where
$\betaPhCE$ is a function of $U$ and $\ket{n}$ is the state with the energy $E_n$.
It may be worth to mention that the condition $n \neq n'$ does not always indicate $E_n \neq E_{n'}$.

We define the following function $J_q(\hA)$:
\begin{align}
J_q(\hA) = \sum_{n}  [\exp_q(-\betaPhCE  \cdot (E_n-U))]^q \bra{n} \hA \ket{n}.
\end{align}
By inserting the delta function, we have
\begin{align}
J_q(\hA) = \sum_{n} \int_{-\infty}^{\infty} dE\ \delta(E-E_n) [\exp_q(-\betaPhCE  \cdot (E-U))]^q \bra{n} \hA \ket{n} .
\end{align}
The state $\ket{j}$ with energy $E_j \neq E$ in the integral does not contribute to $J_q(\hA)$.
We have
\begin{align}
  J_q(\hA) &= \int_{-\infty}^{\infty} dE\
  \sum_{j=1}^{W(E,V,N,\DE)} \delta(E-E_{j})
    [\exp_q(-\betaPhCE  \cdot (E-U))]^q \bra{j} \hA \ket{j}
  \nonumber \\  
  &= \int_{-\infty}^{\infty} dE [\exp_q(-\betaPhCE  \cdot (E-U))]^q \ \Omega_{(EVN)} \ \ExpectMEwithEVN{\hA},
  \label{eqn:JqA}
\end{align}
where we attach the subscript $(E,V,N)$ to show the variables explicitly.
With this function, the escort average $\ExpectCE{\hA}$ is represented by
\begin{align}
  \ExpectCE{\hA} = \frac{J_q(\hA)}{J_q(\One)} .
  \label{CE-ME-relation}
\end{align}
The escort average in the canonical ensemble is represented
with the escort average in the microcanonical ensemble in Eq.~\eqref{CE-ME-relation}.

We can obtain the relation explicitly by calculating $J_q(\hA)$. 
We define the following functions by considering the expansion with respect to $E-U$:
\begin{subequations}
\begin{align}
  &f_q(E) = 1 + \frac{1}{2} q(q-1) (\betaPhCE \cdot (E-U))^2 - \frac{1}{3} q(q-1)^2 (\betaPhCE \cdot (E-U))^3 + O((q-1)^2 (\betaPhCE \cdot (E-U))^4), \\
  &F_q(E) = q \betaPhCE \cdot (E-U) - \ln (\Omega_{(EVN)} f_q(E)) .
\end{align}
\end{subequations}
With the function $F_q(E)$, the  function $J_q(\hA)$ is given by
\begin{align}
J_q(\hA) = \int_{-\infty}^{\infty} dE \ExpectME{\hA} \exp(-F_q(E)) . 
\end{align}
Assuming that $\ExpectME{\hA}$ is a slowly varying function of $E$, we apply the steepest descent method.
We find the extremum $E^*$ of $F_q(E)$.
\begin{align}
\left. \frac{\partial F_q(E)}{\partial E} \right|_{E=E^*} \equiv F_q^{(1)}(E^*) = 0. 
\label{Estar}
\end{align}
The function $J_q(\hA)$ is approximated as
\begin{subequations}
\begin{align}
  J_q(\hA) &\sim \exp(-F_q(E^*)) \int_{-\infty}^{\infty} dE\ \ExpectMEwithEVN{\hA} \exp\left(-\frac{1}{2} F_q^{(2)}(E^*) (E-E^*)^2 \right) \nonumber \\
  &\sim \sqrt{\frac{2\pi}{F_q^{(2)}(E^*)}} \exp(-F_q(E^*)) \left\{
  \left. \ExpectME{\hA}\right|_{E=E^*} + \frac{1}{2F_q^{(2)}(E^*)} \left. \frac{\partial^2 \ExpectME{\hA}}{\partial E^2} \right|_{E=E^*} 
  \right\} \label{expression:JA:Estar} \\
  & = \sqrt{\frac{2\pi}{F_q^{(2)}(E^*)}} \ \Omega_{(E^*VN)} f_q(E^*) \exp(-q \betaPhCE \cdot (E^*-U)) \nonumber \\
  & \qquad \times
    \left\{
    \left. \ExpectME{\hA}\right|_{E=E^*} + \frac{1}{2F_q^{(2)}(E^*)} \left. \frac{\partial^2 \ExpectME{\hA}}{\partial E^2} \right|_{E=E^*} 
    \right\}, 
\end{align}
\end{subequations}
where $F_q^{(2)}(E^*) >0$ is assumed.

We attempt to obtain the solution $E^*$.
The equation \eqref{Estar} is explicitly described:
\begin{align}
F_q^{(1)}(E^*) = q \betaPhCE(U) - \betaPhME(E^*) - \frac{f_q^{(1)}(E^*)}{f_q(E^*)} = 0 , 
\label{steepest-descent}
\end{align}
where $\betaPhME(E)$ is given by $\Omega^{(1)}_{(EVN)}/\Omega_{(EVN)}$. 
We assume the following form of $E^*$ to obtain the solution $E^*$ approximately:
\begin{align}
  E^* = U + (q-1) \lambda,
\label{eqn:app:Estar}
\end{align}
where $\lambda$ is a function of $U$.

We attempt to obtain the quantity $\lambda$.
Equation~\eqref{eqn:app:Estar} is substituted into Eq.~\eqref{steepest-descent},
and we have
\begin{align}
q \betaPhCE(U) - \left(\betaPhME(U) + \left.\betaPhME\right.^{(1)}(U) \cdot (q-1) \lambda \right) + O((q-1)^2) = 0 .
\end{align}
It is natural to assume that $\betaPhCE(U)$ equals $\betaPhME(U)$,
because the inverse of the physical temperature is determined from the R\'enyi entropy:
the R\'enyi entropy in the equilibrium is given as the logarithm of the number of states, $\ln W$. 
The quantity $\lambda$ is given as follows, when this assumption is valid:
\begin{align}
  \lambda = \betaPhCE(U)/\left.\betaPhCE\right.^{(1)}(U) = -\TphCE \CvCE, 
\end{align}
where $\TphCE$ is a function of $U$ 
and $\CvCE$ is the heat capacity at constant volume: $\CvCE$ is given by $({\partial U}/{\partial \TphCE})_V$.

The function $F_q^{(2)}(E)$ is given by
\begin{align}
  F_q^{(2)}(E)
  = \frac{1}{( \TphME(E) )^2} \frac{\partial \TphME(E)}{\partial E}
  - \frac{f_q^{(2)}(E)}{f_q(E)} + \left( \frac{f_q^{(1)}(E)}{f_q(E)} \right)^2 .
\end{align}
Under the assumption $\TphME(U) = \TphCE(U)$,
we obtain
\begin{align}
  F_q^{(2)}(E^*)
  = \frac{1}{\CvCE (\TphCE)^2}
  \Bigg\{ 1 + \Bigg[ 2 - \CvCE + & \left( \frac{\TphCE}{\CvCE} \right)\left( \frac{\partial \CvCE}{\partial \TphCE} \right) \Bigg] (q-1)
  + O((q-1)^2) \Bigg\} .
\end{align}

The condition $F_q^{(2)}(E^*) > 0$ is required for the integral of $J_q(\hA)$ in Eq.~\eqref{expression:JA:Estar}.
The heat capacity $\CvCE$ is positive for standard materials.
This condition indicates that
\begin{align}
1 + \left[ 2 - \CvCE + \left( \frac{\TphCE}{\CvCE} \right)\left( \frac{\partial \CvCE}{\partial \TphCE} \right) \right] (q-1) > 0 .
\label{condition-of-integral}
\end{align}
We note that the quantity $|(\TphCE/\CvCE) (\partial \CvCE/\partial \TphCE)_V|$ may be evaluated 
from the experimental data\cite{Gopal:Book, Miyazaki:Book}.
In the region of $|( {\TphCE}/{\CvCE} )  ({\partial \CvCE}/{\partial \TphCE} ) | \ll 1$,
this condition is reduced to 
\begin{align}
1 + ( 2 - \CvCE ) (q-1) > 0 .
\label{eqn:Fs:condition:reduced}
\end{align}
The above condition is rewritten for $\CvCE \gg 2$: 
\begin{align}
(q-1) \CvCE < 1. 
\label{eqn:condition-by-N}
\end{align}

By expanding the quantity with respect to $(q-1)$ by using Eqs.~\eqref{CE-ME-relation}, \eqref{expression:JA:Estar}, and \eqref{eqn:app:Estar},
we obtain the following relation with the condition, Eq.~\eqref{condition-of-integral}:
\begin{align}
  \ExpectCE{\hA}(U) =&
  \ExpectME{\hA}(U) + \frac{1}{2} \CvCE \cdot (\TphCE)^2 \frac{\partial^2 \ExpectME{\hA}(U)}{\partial U^2} 
   -(q-1) \CvCE \TphCE \Bigg\{
   \frac{\partial \ExpectME{\hA}(U)}{\partial U}
   \nonumber \\
   & + \frac{1}{2} \TphCE \left[
    \CvCE \TphCE \frac{\partial^3 \ExpectME{\hA}(U)}{\partial U^3}
    + \left( 2 - \CvCE + \left(\frac{\TphCE}{\CvCE}\right)\left(\frac{\partial \CvCE}{\partial \TphCE}\right) \right)
      \frac{\partial^2 \ExpectME{\hA}(U)}{\partial U^2}
    \right]
   \Bigg\}  \nonumber \\
   & + O((q-1)^2),
   \label{eqn:CE-ME-relation}
\end{align}
where we note that $\ExpectCE{\hH}$ equals $U$ from the energy constraint.
Equation~\eqref{eqn:CE-ME-relation} is the relation at the energy $U$.
The relation between $\ExpectCE{\hH}$ and $\ExpectME{\hH}$ is discussed in the next section.
In the Boltzmann-Gibbs limit, $q \rightarrow 1$,
Eq.~\eqref{eqn:CE-ME-relation} becomes 
\begin{align}
\left. \ExpectCE{\hA}(U) \right|_{q=1} =
\left. \ExpectME{\hA}(U) \right|_{q=1}
+ \frac{1}{2} \CvCE \cdot (\TphCE)^2 \left. \frac{\partial^2 \ExpectME{\hA}(U)}{\partial U^2} \right|_{q=1} .  
\end{align}

\subsection{The calculation of the partition function}
\label{subsec:partition-function}
The partition function $Z$ was given by Eq.~\eqref{eqn:CE:partition-function}.
We show $Z$ again:
\begin{align}
Z = \Tr{\exp_q( -\betaPhCE  \cdot  (\hH-\ExpectCE{\hH}))}.  
\end{align}
The partition function is rewritten as follows:
\begin{align}
  Z &= \int_{-\infty}^{\infty} dE \ \Omega_{(EVN)} \exp_q( -\betaPhCE  \cdot  (E-U)),
\end{align}
where $\betaPhCE$ is a function of $U$. 
In the similar way, the partition function $Z$ is expressed as 
\begin{subequations}
\begin{align}
Z &= \int_{-\infty}^{\infty} dE \ \exp(-G_q(E)), \\
& G_q(E) = \betaPhCE \cdot (E-U) - \ln \Omega_{(EVN)} - \ln g_q(E) ,\\
& g_q(E) = 1+\frac{1}{2} (q-1) \left( \betaPhCE \right)^2 \cdot (E-U)^2 + O((q-1)^2) .
\end{align}
\end{subequations}
We use the steepest descent method to estimate the partition function $Z$.
The value $E^{\triangledown}$ which satisfies $G^{(1)}(E^{\triangledown}) = 0$ is
\begin{align}
E^{\triangledown} = U + O((q-1)^2) .
\end{align}
The functions $G(E^{\triangledown})$ and $G^{(2)}(E^{\triangledown})$ are given by
\begin{subequations}
\begin{align}
  &G_q(E^{\triangledown}) = - \ln \Omega_{(UVN)} + O((q-1)^2) , \\
  &G_q^{(2)}(E^{\triangledown}) = \frac{1}{\CvCE \cdot \left( \TphCE \right)^2} \left\{ 1 - (q-1) \CvCE \right\} + O((q-1)^2) 
  , \quad \TphCE \equiv \TphCE(U).
\end{align}
\end{subequations}
With these expressions, we estimate the partition function $Z$ for $G_q^{(2)}(E^{\triangledown}) > 0$: 
\begin{align}
  Z &\sim \sqrt{\frac{2\pi}{G_q^{(2)}(E^{\triangledown})}} \exp(-G_q(E^{\triangledown})) . 
\label{expression-Z}
\end{align}

The requirement for the integral $G_q^{(2)}(E^{\triangledown}) > 0$ gives the condition.
For $\CvCE(U) > 0$, this condition is reduced to
\begin{align}
  1 - (q-1) \CvCE > 0.
\label{condition:Cv}
\end{align}
This condition is the generalization of the condition, which is the relation between $q-1$ and the number of independent ingredients $N$,
given in the previous works \cite{Abe-PLA:2001,Ishihara:2022}.
The result in the previous works is recovered for $\CvCE = N$.
It may be worth to mention that Eq.~\eqref{eqn:Fs:condition:reduced} is satisfied for $q>1$ when Eq.~\eqref{condition:Cv} is satisfied.

\section{The quantities related to the energy and the entropies in the Tsallis statistics}
\label{sec:energy-fluctuation-entropies}
\subsection{The quantities related to the energy}
\label{subsec:energy}
We begin with the relation between the energy in the microcanonical ensemble and the energy in the canonical ensemble. 
As is well-known, the energy in the canonical ensemble is equal to the energy in the microcanonical ensemble, in the Boltzmann-Gibbs statistics.
In contrast, from Eq.~\eqref{eqn:CE-ME-relation}, 
the energy in the canonical ensemble is slightly different from the energy in the microcanonical ensemble in the Tsallis statistics.
That is, it is expected that $\ExpectME{\hH}$ differs slightly from $\ExpectCE{\hH}$ which equals $U$.
Therefore, we put $\ExpectME{\hH}$ as follows:
\begin{align}
  \ExpectME{\hH} = U + (q-1) \kappa,
\label{eqn:expansion-H-ME}
\end{align}
where $\kappa$ is a function of $U$. 
By inserting Eq.~\eqref{eqn:expansion-H-ME} into Eq.~\eqref{eqn:CE-ME-relation}, we have 
\begin{align}
\kappa - \CvCE \TphCE + \frac{1}{2} \CvCE \left(\TphCE\right)^2 \frac{\partial^2 \kappa}{\partial U^2} = 0 .
\end{align}
We assume that the first term is sufficiently larger than the third term:
\begin{align}
  \left| \frac{1}{2 \kappa} \CvCE \left(\TphCE\right)^2 \frac{\partial^2 \kappa}{\partial U^2} \right| \ll 1 .
\label{eqn:condition:kappa}
\end{align}
In such the case, we obtain
\begin{align}
\ExpectME{\hH}(U) = U + (q-1) \CvCE \TphCE. 
\label{eqn:ExpectME-H1}
\end{align}

We have the relation between the heat capacities for large $\CvCE$
under the condition that the order of $\TphCE (\partial \CvCE/\partial \TphCE)_V$ is approximately equal to or less than $\CvCE$, 
when $\TphME$ equals $\TphCE$:
\begin{align}
  \CvME = \CvCE + (q-1) \CvCE + (q-1) \TphCE \left(\frac{\partial \CvCE}{\partial \TphCE}\right)_V \sim \CvCE,  
  \label{eqn:CvME-CvCE}
\end{align}
because of the condition $0 < (q-1)\CvCE < 1$.

When the heat capacity $\CvCE$ is approximately $N$ and $U$ is $N\TphCE$,
the energy for large $N$ in the microcanonical ensemble is given by 
\begin{align}
\ExpectME{\hH}(U) \sim  (N + (q-1)N ) \TphCE  \sim N \TphCE,
\label{eqn:HME-HCE}
\end{align}
because $N (q-1)$ is less than one from the condition Eq.~\eqref{condition:Cv}.   
It is found from Eq.~\eqref{eqn:HME-HCE} that
$\ExpectME{\hH}$ is approximately equal to $\ExpectCE{\hH}$ in such the case.

Next, we study the fluctuation of the energy in the canonical ensemble in the Tsallis statistics.
The expectation value of $\hH^2$ in the canonical ensemble differs from that in the microcanonical ensemble, even in the Boltzmann-Gibbs statistics.
The expectation value $\ExpectME{\hH}$ is given by Eq.~\eqref{eqn:ExpectME-H1}.
The energy is constant in the microcanonical ensemble.
Therefore, the expectation value of $\hH^2$ in the microcanonical ensemble is given by
\begin{align}
\ExpectME{\hH^2} = \Big(\ExpectME{\hH}\Big)^2 = U^2 + 2 (q-1) \CvCE \TphCE U + O((q-1)^2). 
\label{eqn:ExpectME-H2}
\end{align}
We note again that $U$ is $\ExpectCE{\hH}$. 
Inserting $\hH^2$ into $\hA$ in Eq.~\eqref{eqn:CE-ME-relation} and using Eq.~\eqref{eqn:ExpectME-H2},
we have
\begin{align}
  &\ExpectCE{\hH^2}(U)
  = U^2 + \CvCE \left( \TphCE \right)^2 + (q-1) \left( \CvCE \TphCE \right)^2 + O((q-1)^2).  
\label{eqn:H-CE}
\end{align}
That is 
\begin{align}
  \ExpectCE{\hH^2}(U) - \left( \ExpectCE{\hH}(U) \right)^2
  =  \CvCE \left( \TphCE \right)^2 \left( 1 + (q-1)  \CvCE \right) + O((q-1)^2).  
\label{Eq:fluc}
\end{align}
In the Boltzmann-Gibbs limit ($q \rightarrow 1$), we have the standard relation in the Boltzmann-Gibbs statistics:
\begin{align}
  \left. \ExpectCE{\hH^2}(U) \right|_{q=1} - \left. \left( \ExpectCE{\hH}(U) \right)^2  \right|_{q=1}
  = \CvCE \Big( \TphCE \Big)^2 . 
\end{align}
From Eq.~\eqref{eqn:H-CE}, we obtain the difference $\ExpectCE{\hH^2}(U) - \ExpectME{\hH^2}(U)$:
\begin{align}
  \ExpectCE{\hH^2}(U) - \ExpectME{\hH^2}(U) 
  &=  \CvCE \left( \TphCE \right)^2 \left[ 1 + (q-1)  \left( \CvCE - \frac{2U}{\TphCE} \right) \right] + O((q-1)^2) .
  \label{Eq:fluc:diff}
\end{align}
The quantity $1 + (q-1)  ( \CvCE - {2U}/{\TphCE})$ in Eq.~\eqref{Eq:fluc:diff} equals $1-N(q-1)$ in the case of $U=N \TphCE$.
The inequality $N (q-1) >0$ is satisfied for $q>1$ and the constraint $1 - (q-1)\CvCE > 0$ equals $1-N(q-1) >0$. 
It is easily found in such the case that $\ExpectCE{\hH^2} - \ExpectME{\hH^2}$ is positive.

\subsection{The quantities related to the R\'enyi and the Tsallis entropies}
The quantity $W(E,V,N,\DE)$ is represented as $W(E,V,N,\DE)=\Omega_{(E.V,N)} \DE $ for small $\DE$.
The R\'enyi entropy in the microcanonical ensemble is 
\begin{align}
\SRME = \ln (W(E,V,N,\DE)) = \ln (\Omega_{(E,V,N)} (\DE)).
\end{align}

With Eqs.~\eqref{expression:JA:Estar} and \eqref{expression-Z}, 
the R\'enyi entropy in the canonical ensemble is given by
\begin{align}
  \SRCE
  &= \frac{1}{1-q} \ln \left( \Tr{\left(\qmrhoCE\right)^q} \right) 
  = \frac{1}{1-q} \ln \left( Z^{-q} J_q(\One) \right)
  = \frac{1}{1-q} \left\{ \ln J_q(\One) - \ln Z - (q-1) \ln Z \right\}
  \nonumber \\
  &= \SRME + \ln (\TphCE/(\DE))  + \frac{1}{2} \ln \CvCE + 1 + \frac{1}{2} \ln(2\pi)
  + \frac{1}{2} \left(\frac{\TphCE}{\CvCE}\right) \left(\frac{\partial\CvCE}{\partial\TphCE}\right) + O(q-1) .
\label{rel:SRCE-SRME}
\end{align}

The magnitude of the relative difference, $|(\SRCE-\SRME)/\SRME|$ can be estimated. 
The term $\ln(\CvCE)$ is probably the leading term for sufficiently large $\CvCE$. 
The quantity $\SRCE-\SRME$ is approximately given by $\ln(\CvCE)/2$ in such the case.
For sufficiently large $\CvCE$, the magnitude of the relative difference is estimated as
\begin{align}
\left| \frac{\SRCE-\SRME}{\SRME} \right| \sim  \frac{1}{2} \frac{\ln(\CvCE)}{\ln(\Omega_{(EVN)}(\DE))}.  
\end{align}
We must pay attention to the condition $\CvCE (q-1) < 1$.

As an example, we treat the following case \cite{Ishihara:2022,Shirts:2021}:
\begin{align}
  &\Omega_0(E,V,N) = \eta \left(\frac{E}{N}\right)^{\alpha N}, 
  \qquad \eta = \eta(V,N) ,
\label{special-case}
\end{align}
where $V$ is the volume and  $N$ is the number of ingredients.
For free particles of mass $m$ with $\alpha=3/2$, the quantity $\eta$ is
\begin{align}
\eta(V,N) = \frac{V^N}{h^{3N} N!} \frac{(2\pi m)^{\frac{3}{2}N} }{\Gamma\left(\frac{3}{2} N + 1\right)} N^{\frac{3}{2} N},  
\end{align}
where $h$ is the Planck's constant and $\Gamma(x)$ is the Gamma function. 
The magnitude of the relative difference is estimated for sufficiently large $N$ in the system of $N$ free particles when $V/N$ is (approximately) constant:
\begin{align}
  \left| \frac{\SRCE-\SRME}{\SRME} \right| = O(N^{-1} \ln N) .
  \label{ratio:SRCE-SRME}
\end{align}
Therefore, the difference $\SRCE-\SRME$ is small compared with $\SRME$ for large $N$.
The relative difference approaches zero as $N$ approaches infinity.
It is noted that the value of $N$ is restricted through the condition between the heat capacity and the entropic parameter. 

The relative difference in the Tsallis entropy is estimated through the relative difference in the R\'enyi entropy:
\begin{align}
\frac{\STCE-\STME}{\STME} = \frac{1-\exp\left(-(q-1)\SRME \left(\frac{\SRCE-\SRME}{\SRME}\right)\right)}{\exp((q-1)\SRME)-1}. 
\label{Tsallis:relative-diff}
\end{align}
The quantity $\SRME$ is the order of $N$ for free particles.
Therefore, the quantity  $(q-1) \SRME$ is the order of one when the inequality $0 < (q-1) \CvCE < 1$ is satisfied, 
because $\CvCE$ is the order of $N$.
With Eq.~\eqref{ratio:SRCE-SRME}, the numerator of Eq.~\eqref{Tsallis:relative-diff} for large $N$ is
\begin{align}
  &1-\exp \left( -(q-1) \SRME \left(\frac{\SRCE-\SRME}{\SRME}\right) \right)
  \nonumber \\
  &= (q-1) \SRME \left(\frac{\SRCE-\SRME}{\SRME}\right)
  + O \left( \left[ (q-1)\SRME \left(\frac{\SRCE-\SRME}{\SRME}\right) \right]^2 \right) . 
\end{align}
We have
\begin{align}
  \frac{\STCE-\STME}{\STME} \sim \frac{(q-1)\SRME}{\exp((q-1)\SRME)-1} \left(\frac{\SRCE-\SRME}{\SRME}\right).
\end{align}
Therefore, the order of the relative difference $(\STCE-\STME)/\STME$ is $O(N^{-1}\ln N)$ at most,
because the inequality $0< x/(\exp(x)-1) \le 1$ holds for $x \ge 0$. 

\section{Discussions and Conclusions} 
We studied the escort averages in microcanonical and canonical ensembles in the Tsallis statistics of the entropic parameter $q>1$.
The Boltzmann-Gibbs statistics corresponds to the Tsallis statistics of $q=1$. 
We obtained the relation between the escort average in the microcanonical ensemble and the escort average in the canonical ensemble.
We investigated the relations for the energy, the expectation value of $\hH^2$, the R\'enyi entropy, and the Tsallis entropy,   
where $\hH$ is the Hamiltonian.

The relation between the escort average in the microcanonical ensemble and the escort average in the canonical ensemble
in the Tsallis statistics is different from that in the Boltzmann-Gibbs statistics.
The relation in the Boltzmann-Gibbs statistics is recovered in the limit $q \rightarrow 1$.
The relation is influenced by the statistics, and
the quantity in the canonical ensemble is not always equal to the quantity in the microcanonical ensemble.
This relation will be used to derive the quantities such as the fluctuation, as shown in this paper.

The condition between the value $q-1$ and the heat capacity $\CvCE$ at constant volume in the canonical ensemble arises
from the requirement that the integrals appeared in the canonical ensemble do not diverge.
The existence of the condition indicates the importance of the heat capacity in the Tsallis statistics.
The difference between the quantity in the microcanonical ensemble and the quantity in the canonical ensemble is affected
by the condition: $0<(q-1) \CvCE < 1$.
In the case that $\CvCE$ is approximately the number of ingredients $N$, we obtain the condition $(q-1)N<1$ which was found in the previous works. 

The energy $\ExpectCE{H}$ in the canonical ensemble differs slightly the energy $\ExpectME{H}$ in the microcanonical ensemble.
Using the relation between $\ExpectME{H}$ and $\ExpectCE{H}$, 
we may rewrite the quantity as a function of $\ExpectCE{H}$ with $\ExpectME{H}$.
The difference between $\ExpectME{H}$ and $\ExpectCE{H}$ is small when the condition $0 < (q-1) \CvCE < 1$ holds.
Therefore, $\ExpectCE{H}$ is approximately equal to $\ExpectME{H}$ for large $N$ in the case that $\ExpectCE{H}$ is represented as $N\TphCE$,
where $\TphCE$ is the physical temperature in the canonical ensemble. 
These energies take the same value in the Boltzmann-Gibbs limit, $q \rightarrow 1$, as is well-known.

The energy fluctuation is affected by the distribution.  
The energy fluctuation in canonical ensemble, $\ExpectCE{\hH^2} - (\ExpectCE{\hH})^2$, is modified
with additional term which is proportional to $(q-1)$. 
Therefore, the relation between the fluctuation and the heat capacity
approaches the well-known relation in the Boltzmann-Gibbs statistics as $q$ approaches one.
The difference between the canonical ensemble and the microcanonical ensemble in the expectation value of $\hH^2$ is also modified.
This difference depends on the energy, temperature, and heat capacity.

The partition function was also estimated, 
and the condition between the quantity $(q-1)$ and the heat capacity $\CvCE$ in the canonical ensemble appears.
The condition is simple: $1-(q-1) \CvCE >0$.
It is evident that the value $q$ must be one when the heat capacity is infinite.
This condition generate $1-(q-1) N>0$ in the case that $\CvCE$ is equal to $N$. 

The magnitude of the relative difference  $\left| (\SRCE-\SRME)/\SRME \right|$, was estimated, 
where $\SRCE$ is the R\'enyi entropy in the canonical ensemble and $\SRME$ is the R\'enyi entropy in the microcanonical ensemble.
The magnitude of the relative difference is the order of $N^{-1} \ln N$ for free particles of mass $m$:
the number of states below the energy $E$ is proportional to $E^{3N/2}$. 
As expected, the relative difference in the R\'enyi entropy is small for large $N$.
We note that the quantity $N$ is restricted because of the condition, $0 < (q-1) \CvCE < 1$.
The quantity $\left| (\STCE-\STME)/\STME \right|$ was estimated
by using the relation between the Tsallis and the R\'enyi entropies, 
where $\STCE$ is the Tsallis entropy in the canonical ensemble and $\STME$ is the Tsallis entropy in the microcanonical ensemble.
The relative difference in the Tsallis entropy is small for large $N$ like the relative difference in the R\'enyi entropy.

The heat capacity has an important role in the Tsallis statistics:
the Tsallis distribution was derived in the system with constant heat capacity \cite{Wada2003}
and the relation between the entropic parameter and the heat capacity was found in the system of fluctuating temperature \cite{Wilk:EPJA40:2009}.
As shown in this paper,
the condition between the entropic parameter and heat capacity 
affects the relation between the escort average in the microcanonical ensemble and the escort average in the canonical ensemble. 

In this paper,
we found the relation between the escort average in the microcanonical ensemble and the escort average in the canonical ensemble
in the Tsallis statistics.
We can apply this equation to physical quantities, and will find the relation between them.
I hope that this work is useful to find the differences between the Boltzmann-Gibbs statistics and the unconventional statistics
like the Tsallis statistics.

\medskip\noindent
\textbf{Funding}
This research received no specific grant from any funding agency in the public, commercial, or not-for-profit sectors.

\medskip\noindent
\textbf{Data availability}
This manuscript has no associated data or the data will not be deposited. [Authors' comment: This study is theoretical, and no data is generated].

\medskip\noindent
\textbf{Conflict of Interest}
The author declares no competing interest.


\end{document}